\documentclass[12pt]{iopart}
\pdfoutput=1
\usepackage{graphicx}
\usepackage{mathtools}
\usepackage{todonotes}
\usepackage[normalem]{ulem}
\usepackage{xcolor}
\usepackage{caption}
\usepackage{subcaption}
\usepackage{textcomp}
\usepackage{float}
\usepackage{soul,comment}
\usepackage{stackrel}
\usepackage{hyperref}
\usepackage{placeins}

\definecolor{dgreen}{rgb}{0,0.7,0}

 \expandafter\let\csname equation*\endcsname\relax
 \expandafter\let\csname endequation*\endcsname\relax

\usepackage{amsmath,amssymb}
\usepackage{soul}

\pdfoutput=1
\usepackage{esint}
\usepackage{amsfonts}
\usepackage{color}
\definecolor{dgreen}{rgb}{0,0.7,0}

\usepackage[
    backend=biber,
    doi=false,
    isbn=false,
    url=false,
    eprint=false,
    style=nature,
    sorting = none
]{biblatex}
\begin{document}

\title[]{
Mpemba effect in a chemomechanical model of the Kinesin molecular motor}

\author{Karthik Cheruvary and Arnab Pal}
\address{The Institute of Mathematical Sciences, CIT Campus, Taramani, Chennai 600113, India \& Homi Bhabha National Institute, Training School Complex, Anushakti Nagar, Mumbai 400094, India}
\ead{kar1504thik@gmail.com,arnabpal@imsc.res.in}
\vspace{5pt}
 
\begin{abstract}
The Mpemba effect, wherein a system prepared farther from equilibrium relaxes faster than one initially closer to equilibrium, has been extensively investigated in a wide range of physical systems. In contrast, its role in biologically relevant non-equilibrium processes remains largely unexplored. Here, we investigate anomalous relaxation in the six-state chemomechanical network model of the Kinesin molecular motor under both equilibrium and non-equilibrium conditions. We first establish the existence of the Mpemba effect in chemical equilibrium and show that many of its qualitative features can be understood from the underlying free-energy landscape. We then examine the effects of mechanical and chemical driving, showing that breaking detailed balance primarily reshapes the Mpemba phase diagram without qualitatively altering the relaxation phenomenology over the physically relevant parameter regime. Finally, we demonstrate that the relaxation of the motor velocity also mirrors the anomalous relaxation of the underlying stochastic dynamics, thereby identifying an experimentally accessible signature of the Mpemba effect. Our results establish molecular motors as a promising baseline for studying anomalous relaxation in living systems and suggest a broader framework for exploring the Mpemba effect in non-equilibrium biochemical networks.
\end{abstract}

\section{Introduction}\label{section_intro}
The Mpemba effect refers to the anomalous dependence of post-quench relaxation on the initial state of a system, whereby a system prepared farther from equilibrium may relax faster than one prepared closer to equilibrium. It was originally reported in the context of water freezing, where a sample initially at a higher temperature was observed to freeze before one at a lower temperature~\cite{mpembaCool1969}. Although the original experiments remain controversial~\cite{burridgeQuestioningMpembaEffect2016}, analogous anomalous relaxation has since been identified in a broad range of equilibrium and non-equilibrium systems, demonstrating that relaxation need not proceed in a sequential manner determined solely by the initial distance from the steady state~\cite{luNonequilibriumThermodynamicsMarkovian2017,biswasMpembaEffectLangevin2023,biswasMpembaEffectDriven2020,lasantaWhenHotterCools2017,santosMpembaEffectMolecular2020,torrenteLargeMpembalikeEffect2019,baity-jesiMpembaEffectSpin2019,chatterjeeMpembaEffectPure2024,ghoshSimulationsMpembaEffect2025,antonovTemperatureOvershootingMpemba2026,mellesQuantizationClassicalMpemba2026,lapollaFasterUphillRelaxation2020,takadaMpembaEffectInertial2021}. This ubiquity has established the Mpemba effect as a paradigmatic example of anomalous relaxation in non-equilibrium statistical physics, with potential implications for accelerated equilibration and optimal control of relaxation processes. 


Despite this progress, a general understanding of the Mpemba effect remains elusive. No universal criterion capable of predicting its occurrence from the steady-state properties of a system has yet emerged. Proposed explanations based on metastable states, confining boundaries, or the spectral properties of the relaxation operator each account for the phenomenon only in specific classes of models~\cite{biswasMpembaEffectLangevin2023,walkerAnomalousThermalRelaxation2021,biswasMpembaEffectRelaxation2025,liuMpembaEffectLikes2026,liuPredictingConditionsObserving2026,biswasMpembaEffectNonequilibrium2025,hayakawaMpembaEffectTwodimensional2026}. Experimentally, the phenomenon has been investigated primarily in controlled colloidal systems involving optically trapped Brownian particles~\cite{kumarExponentiallyFasterCooling2020,kumarAnomalousHeatingColloidal2022,bechhoeferFreshUnderstandingMpemba2021,chetriteMetastableMpembaEffect2021}. A comprehensive overview of recent developments is provided in Ref.~\cite{tezaSpeedupsNonequilibriumThermal2026}. In contrast, the possibility of anomalous relaxation in biological systems remains almost completely unexplored. This is particularly surprising because many intracellular processes are naturally described as stochastic transitions on kinetic networks operating far from equilibrium. Such systems provide a fundamentally different setting from the physical and chemical systems studied so far, where non-equilibrium driving is sustained by continuous energy consumption. Demonstrating the Mpemba effect in this context would therefore not only extend its domain of applicability, but also establish anomalous relaxation as a generic feature of relaxation in living non-equilibrium matter.

Among biological systems, molecular motors provide a particularly attractive platform for such an investigation. Kinesins are ATP-driven motor proteins responsible for intracellular cargo transport, cytoskeletal organization, and mitotic spindle formation. Their stepping motion arises from successive binding and unbinding of the motor domains to a microtubule, powered by ATP hydrolysis, thereby functioning as chemomechanical cyclic engines that convert chemical free energy into directed mechanical motion~\cite{yildizMechanismRegulationKinesin2025,howardMechanicsMotorProteins2024,hirokawaKinesinSuperfamilyMotor2009}. Owing to decades of single-molecule experiments and theoretical developments, Kinesin dynamics is now quantitatively understood through kinetic network models that faithfully capture the underlying chemomechanical cycle~\cite{lipowskyChemomechanicalCouplingMolecular2008}. These models provide a natural and experimentally relevant framework for studying relaxation in living non-equilibrium systems.

In this work, we initiate the study of the Mpemba effect in living systems by investigating the six-state Markov network model of Kinesin studied by Lipowsky and co-authors in Refs.~\cite{liepeltKinesinsNetworkChemomechanical2007,liepeltSteadystateBalanceConditions2007,lipowskyChemomechanicalCouplingMolecular2008}. The structure of the paper follow. We first discuss the network describing Kinesin in Sec.\ref{section_model}, stating all model assumptions and parameters needed to specify its dynamics. The general definition of the Mpemba effect in terms of a crossing in distance measures and the spectral criterion used to detect its presence are then discussed in Sec.\ref{section_mpemba}. The corresponding energy landscape picture is depicted in Sec. \ref{section_energy_landscape}. Sec.\ref{section_results_eq} establishes the existence of the Mpemba effect under equilibrium conditions, corresponding to chemical equilibrium of ATP hydrolysis in the absence of an external load. We illustrate the dependence of the Mpemba phase on system parameters and provide a qualitative explanation based on the sytems underlying free energy landscape. In Sec.\ref{section_results_non_eq} we then investigate how it evolves as chemical and mechanical non-equilibrium are introduced independently. We relate the observed anomalous relaxation to modifications in the underlying free-energy landscape of the kinetic network. Finally, in Sec. \ref{section_results_current}, we identify the motor velocity as an experimentally accessible observable for characterizing the Mpemba effect. We conclude with a summary and future outlook.

\section{Kinetic network model of Kinesin}\label{section_model}

To investigate the Mpemba effect in a biologically relevant setting, we employ the six-state kinetic network model for Kinesin introduced by Lipowsky \textit{et al.}~\cite{liepeltSteadystateBalanceConditions2007,liepeltKinesinsNetworkChemomechanical2007,lipowskyChemomechanicalCouplingMolecular2008}. The model captures the essential chemomechanical features of the molecular motor while remaining sufficiently simple to permit a detailed analysis of its relaxation dynamics.

\begin{figure}[htb!]
    \centering
    \includegraphics[width =\textwidth]{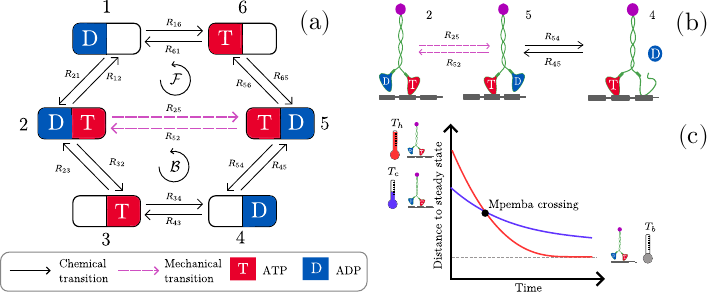}
    \caption{Panel (a): The chemo mechanical network representing Kinesin motion. This numbering of states will be used throughout. $\mathcal{F}$ is the forward cycle and $\mathcal{B}$ is the backward cycle, directed along the arrows. Panel (b): Schematic of Kinesin walker on a microtubule undergoing mechanical and chemical transitions. Panel (c) shows that the Kinesin system initially equilibrated at a higher temperature relaxes faster than the same initially equilibrated at a lower temperature, yielding Mpemba crossing.} 
    \label{Fig_MC_network}
\end{figure}

A Kinesin molecule consists of two motor domains (heads), each possessing binding sites for ATP and the microtubule. The two heads are connected through flexible neck linkers to a common stalk, whose opposite end binds to the cargo. During motion, the motor advances along a microtubule through the characteristic ``hand-over-hand'' stepping mechanism, illustrated schematically in Fig.~\ref{Fig_MC_network}(b).

The mechanochemical state of the motor is determined by the nucleotide occupancy of its two heads. Each head can be empty, ATP-bound, or ADP-bound, giving rise to nine possible configurations connected by 36 possible transitions. However, physical and biochemical constraints eliminate many of these states and transitions, reducing the dynamics to the six-state bicyclic network shown in Fig.~\ref{Fig_MC_network}(a). 

\FloatBarrier
\FloatBarrier

Throughout this work, the state of the motor is represented by an ordered pair specifying the nucleotide states of the lagging and leading heads, respectively. Solid edges denote chemical transitions involving ATP, ADP, or phosphate binding and release, whereas the dashed edge represents the mechanical stepping transition that exchanges the leading and lagging heads. In the absence of an external load, this transition corresponds to the preferred direction of motion of the motor.

The operation of the network is conveniently understood through the forward chemomechanical cycle $\mathcal{F}$. ADP-bound motor domains possess weak affinity for the microtubule. Starting from the state in which both heads are ADP-bound, one head releases ADP and binds to the microtubule, leading to state 1. ATP binding at the leading head ($1\rightarrow2$) induces conformational changes in the neck linker that propel the trailing head forward, giving rise to the mechanical stepping transition ($2\rightarrow5$). Subsequent nucleotide exchange, ATP hydrolysis, and detachment of the lagging head complete the cycle, returning the motor to state 1. Each completion of the forward cycle consumes one ATP molecule and advances the motor by one step. The reverse cycle, denoted by $\mathcal{B}$, likewise hydrolyzes one ATP molecule but produces a backward step~\cite{yildizMechanismRegulationKinesin2025,liepeltSteadystateBalanceConditions2007}.

We assume that the motor is immersed in a thermal bath at temperature $T_b$ and coupled to a chemical reservoir with fixed concentrations of ATP, ADP, and inorganic phosphate (P). Since the state space explicitly resolves only the nucleotide occupancy of the motor domains, all faster internal conformational fluctuations are assumed to equilibrate between successive interstate transitions, analogous to the separation of time scales underlying Kramers' theory. Chemical equilibrium corresponds to ATP hydrolysis being in equilibrium with the surrounding chemical bath,
\begin{equation}
    K_{eq}=\frac{[ADP][P]}{[ATP]}.
    \label{Eq_K_eq}
\end{equation}
The transition rate from state $i$ to state $j$ is written as
\begin{equation}
    R_{ij}=\kappa_{ij}\mathcal{I}_{ij}\phi_{ij}(F),
    \label{Eq_Rij}
\end{equation}
where the three factors represent distinct physical mechanisms. The intrinsic rate $\kappa_{ij}$ accounts for thermal activation over the energy barrier separating the two states,
\begin{align}
    \kappa_{ij}
    =
    \exp\left[-\frac{B_{ij}-E_i}{k_bT}\right],
    \qquad
    i\neq j.
    \label{Eq_intrinsic_rate}
\end{align}
Here $E_i$ denotes the free energy of state $i$, while $B_{ij}$ is the corresponding transition-state barrier. The factor $\mathcal{I}_{ij}$ accounts for coupling to the chemical reservoir. If the transition involves binding of a chemical species $X$, then $\mathcal{I}_{ij}=[X]$; otherwise $\mathcal{I}_{ij}=1$. Finally, $\phi_{ij}(F)$ describes the influence of an external load force $F$. We assume that only the mechanical stepping transition depends explicitly on the applied load,
\begin{equation}
    \phi_{25}=e^{-\theta F},
    \qquad
    \phi_{52}=e^{(1-\theta)F},
    \label{Eq_force_terms}
\end{equation}
with all remaining transitions satisfying $\phi_{ij}=1$. Consequently, a positive load opposes forward stepping of the motor. In reality, the external force modifies the complete free-energy landscape and therefore influences all transition rates. Here we retain only its dominant effect on the mechanical step, thereby keeping the number of independent model parameters to a minimum.

In the absence of an external load and under chemical equilibrium, the kinetic network satisfies detailed balance,
\begin{equation}   \mathbf{P}^{ss}_iR_{ij}=\mathbf{P}^{ss}_jR_{ji},
\end{equation}
where $\mathbf{P}^{ss}_i$ denotes the steady-state occupation probability of state $i$. Consequently, for any closed cycle $\mathcal{C}$ in \textit{equilibrium}, 
\begin{equation}
     \prod_{i\to j \in\mathcal{C}}\frac{\mathbf{P}^{ss}_iR_{ij}}{\mathbf{P}^{ss}_jR_{ji}}
     =
     \prod_{i\to j \in\mathcal{C}}\frac{R_{ij}}{R_{ji}}
     =
     1,
     \label{Eq_cyclic_affinity}
\end{equation}
Since $R_{ij}=\kappa_{ij}\mathcal{I}_{ij}$ when $F=0$, applying Eq.~(\ref{Eq_cyclic_affinity}) to the forward and backward cycles yields
\begin{align}
    1 &=
    \frac{\kappa_{12}[T]\kappa_{25}\kappa_{56}\kappa_{61}}
    {\kappa_{21}\kappa_{52}\kappa_{65}[D]\kappa_{16}[P]}
    =
    \frac{\kappa_{23}\kappa_{34}\kappa_{45}[T]\kappa_{52}}
    {\kappa_{32}[D]\kappa_{43}[P]\kappa_{54}[D]\kappa_{25}},
    \label{dbalance1}\\
    K_{eq}
    &=
    \frac{\kappa_{12}\kappa_{25}\kappa_{56}\kappa_{61}}
    {\kappa_{21}\kappa_{52}\kappa_{65}\kappa_{16}}
    =
    \frac{\kappa_{23}\kappa_{34}\kappa_{45}\kappa_{52}}
    {\kappa_{32}\kappa_{43}\kappa_{54}\kappa_{25}}.
    \label{Eq_detailed_balance}
\end{align}
The intrinsic rates are constructed from a consistent energy landscape and therefore satisfy detailed balance by construction. Consequently, Eq.~(\ref{Eq_detailed_balance}) implies $K_{eq}=1$, corresponding to chemical equilibrium between ATP, ADP, and phosphate. Introducing an external load breaks this equilibrium through the asymmetric force dependence of the mechanical transition rates, $\phi_{25}$ and $\phi_{52}$, thereby violating Eq.~(\ref{Eq_cyclic_affinity}). Thus, a non-zero load drives the motor into a non-equilibrium steady state.

The transition rates are parameterized as follows. We first specify the free energies of the six states together with the transition-state barriers along the forward cycle. The ATP and ADP concentrations are treated as independent parameters, while under chemical equilibrium the phosphate concentration is fixed by the condition $K_{eq}=1$. For the backward cycle, all intrinsic rates are taken to be identical to those of the corresponding forward transitions, except for the transition $5\rightarrow4$. Specifically, we choose
$
\kappa_{12}=\kappa_{45},
\kappa_{61}=\kappa_{34},$
and similarly for the remaining corresponding transitions. The rate $\kappa_{54}$ is then determined by enforcing detailed balance for the backward cycle. Eliminating the common factors between the forward and backward cycles in Eq.~(\ref{Eq_detailed_balance}) gives
\begin{equation}
    \kappa_{54}
    =
    \left(\frac{\kappa_{52}}{\kappa_{25}}\right)^2
    \kappa_{21}.
    \label{Eq_constrained_rate_const}
\end{equation}
The choice of $\kappa_{54}$ as the dependent rate is arbitrary; any one of the backward-cycle rates could instead be determined from the detailed balance constraint.

Chemical non-equilibrium is introduced by allowing the ATP hydrolysis reaction to deviate from equilibrium while leaving the intrinsic energy landscape unchanged. We therefore continue to fix the ATP and ADP concentrations, but determine the phosphate concentration according to
\begin{equation}
    [P]
    =
    c \cdot K_{eq}\frac{[ADP]}{[ATP]},
    \label{Eq_P_conc}
\end{equation}
where the dimensionless parameter $c$ quantifies the degree of departure from chemical equilibrium. For $c=1$, the chemical bath is in equilibrium, whereas $c\neq1$ corresponds to non-equilibrium chemical driving. In this case Eq.~(\ref{dbalance1}) is no longer satisfied because detailed balance is broken by the chemical reservoir. However, Eq.~(\ref{Eq_detailed_balance}) continues to hold since the intrinsic rates remain unchanged. Accordingly, $\kappa_{54}$ is specified exactly as in the equilibrium case.

With the transition rates completely specified, the time evolution of the occupation probabilities is governed by the corresponding master equation, discussed in the following section. These time-dependent probabilities constitute the primary quantities used to characterize relaxation and thereby investigate the Mpemba effect.

In addition to the state occupation probabilities, we also study the relaxation of the motor velocity, which provides an experimentally accessible observable for characterizing the Mpemba effect. If each mechanical step advances the motor by a distance $l$, the instantaneous velocity is determined by the probability current across the mechanical transition,
\begin{equation}
    v(t)=lJ_{25}(t)
    =
    l[\mathbf{P}_2(t)R_{25}-\mathbf{P}_5(t)R_{52}],
    \label{Eqn_J25}
\end{equation}
where $J_{25}(t)$ denotes the net probability current from state 2 to state 5 and $\mathbf{P}_i(t)$ denotes the occupation probability of state $i$. Since the mechanical transition is responsible for forward stepping, the motor velocity directly reflects the competition between forward and backward transport pathways.

\FloatBarrier
\begin{figure}[htb!]
    \centering
    \includegraphics[width = \textwidth]{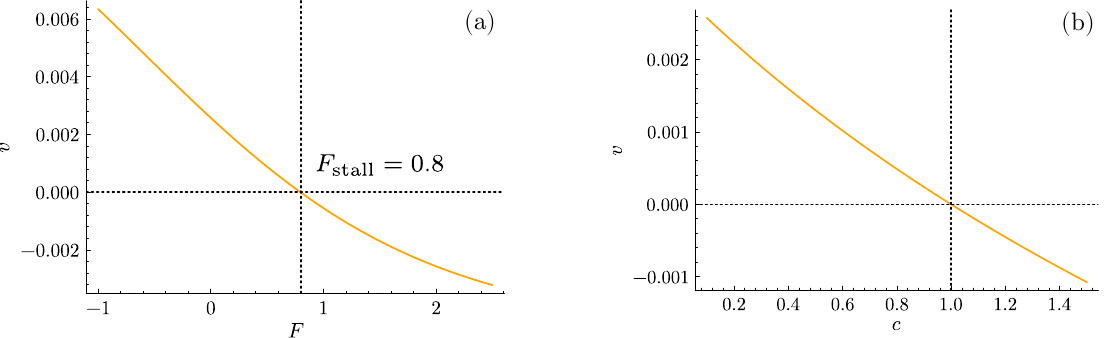}
    \caption{Non-equilibrium nature of the Kinesin depicted by the velocity defined via the current. Panel (a): Steady state velocity $v$ vs load force $F$, showcasing $v$ decreasing with increasing load and stalling at $F_{\text{stall}} = 0.8$. Other parameters are $T_b = 1, s = 0.5, r = 0.1, c = 0.1, \theta = 0.6$. Since $c\neq1$, the network is out of equilibrium. Panel (b): Velocity $v$ vs chemical driving $c$ for $T_b = 1, s = 0.5, r = 0.1, F = 0$. $c = 1$ corresponds to chemical equilibrium, resulting in $v = 0$.} 
    \label{Fig_velocity_entropy_prod_vs_F}
\end{figure}
\FloatBarrier

As shown in Fig.~\ref{Fig_velocity_entropy_prod_vs_F}, varying either the external load or the chemical driving parameter $c$ generates finite velocity, signaling the emergence of nonequilibrium steady states. As the load force is increased, the motor velocity reduces and it stalls when the net current vanishes ($v=0$). The characteristic force at which this happens is called the stall force $F_{\text{stall}}$. For loads exceeding the stall force, the external force drives the motor in the backward direction, despite the continued consumption of ATP. Within the bicyclic network, this seemingly counterintuitive behavior is naturally explained by the competition between the forward and backward chemomechanical cycles, both of which hydrolyze ATP while generating motion in opposite directions~\cite{lipowskyChemomechanicalCouplingMolecular2008,liepeltKinesinsNetworkChemomechanical2007,liepeltSteadystateBalanceConditions2007}. For sufficiently large loads, additional mechanisms such as slippage become important~\cite{carterMechanicsKinesinStep2005}. Since these processes are not incorporated in the present model, our analysis is restricted to load regimes where the six-state description remains valid.

Our primary objective is to investigate transient relaxation rather than quantitatively reproduce the steady-state transport characteristics of Kinesin. Accordingly, the intrinsic rates are chosen to explore a broad range of physically consistent energy landscapes while preserving the qualitative behavior of the motor. In the following section, we formulate the corresponding master equation and analyze how these non-equilibrium conditions influence the relaxation dynamics and the emergence of the Mpemba effect.

\section{Relaxation dynamics and the Mpemba effect}\label{section_mpemba}

Having specified all transition rates, we now turn to the relaxation dynamics of the motor. The time evolution of the occupation probabilities is governed by the continuous-time master equation. The corresponding transition-rate matrix is given by
\begin{equation}
    M_{ij}=
    \begin{cases}
        R_{ji}, & i\neq j,\\
        -\displaystyle\sum_{k\neq i}R_{ik}, & i=j,
    \end{cases}
    \label{Eq_rate_matrix}
\end{equation}
where $R_{ij}$ are the transition rates defined in the previous section. The probability vector $\mathbf{P}(t)$ therefore evolves according to
\begin{equation}
    \frac{d\mathbf{P}}{dt}
    =
    M\mathbf{P},
    \label{Eq_master_eqn}
\end{equation}
Since the transition rates depend on the bath temperature, chemical concentrations, and external load, the generator $M$ is itself a function of these control parameters.

The Mpemba effect is traditionally studied by considering temperature quenches. Two identical copies of the system are initially prepared in their respective steady states at temperatures $T_h$ and $T_c$, with $T_h>T_c>T_b$. Both systems are then instantaneously quenched to the common bath temperature $T_b$, and their subsequent relaxation towards the new steady state is monitored. The Mpemba effect is said to occur when the system prepared initially at the higher temperature approaches the final steady state faster than the one prepared at the lower temperature.

\subsection{Quantifying relaxation}\label{section_mpemba_def}

The Mpemba effect has been argued to be independent of the particular measure used to quantify the distance from the steady state, provided the measure satisfies three natural conditions:\cite{luNonequilibriumThermodynamicsMarkovian2017}

\begin{enumerate}
    \item $D[\mathbf{P}^{ss}(T_h),\mathbf{P}^{ss}(T_b)] >
    D[\mathbf{P}^{ss}(T_c),\mathbf{P}^{ss}(T_b)]$,
    \item $D[\mathbf{P}_T(t),\mathbf{P}^{ss}(T_b)]$ is a non-increasing function of time,
    \item $D[\mathbf{P}(t),\mathbf{P}^{ss}(T_b)]$ is a convex function of $\mathbf{P}(t)$.
\end{enumerate}

Here, $\mathbf{P}^{ss}(T)$ denotes the steady-state distribution at bath temperature $T$, while $\mathbf{P}_T(t)$ is the probability distribution obtained after quenching the steady state at temperature $T$ to the final bath temperature $T_b$. Condition (i) is naturally satisfied for temperature quenches, since increasing the initial temperature generally increases the distance from the final steady state. However, this need not hold for other control parameters, such as the mechanical quench considered later in this work. In those cases, we regard a monotonic increase of the initial distance with the quenched parameter as a prerequisite for discussing the Mpemba effect.

Several commonly used measures, including the entropic distance ($L_1$) and the Kullback--Leibler divergence, satisfy the remaining conditions. Throughout this work we employ the $L_1$ distance,
\begin{align}
    |\mathbf{P}-\mathbf{Q}|_{L_1}
    &=
    \sum_{i=1}^{n}
    |\mathbf{P}_i-\mathbf{Q}_i|,
    \label{Eq_l1_norm}\\
    L^T_1(t)
    &\coloneqq
    |\mathbf{P}_T(t)-\mathbf{P}^{ss}(T_b)|_{L_1},
    \label{Eq_mpemba_dist}
\end{align}
where, for notational simplicity, the dependence on the initial and final temperatures is suppressed whenever no ambiguity arises. Thus, the Mpemba effect is said to occur if $L^{T_h}_1(t) < L^{T_c}_1(t)$ for all $t$ greater than some crossing time $t_c$, provided $L^{T_h}_1(0) > L^{T_c}_1(0)$ \cite{luNonequilibriumThermodynamicsMarkovian2017}. A representative example of the resulting crossing of the relaxation curves is shown in Fig.~\ref{Fig_mpemba_example}(a). A set of more restrictive conditions that ensures a bounded set of crossing times over all suitable distance functions is discussed in Ref. \cite{vanvuThermomajorizationMpembaEffect2025}.

\subsection{Spectral criterion for the Mpemba effect}\label{section_mpemba_a2}

For Markovian systems satisfying detailed balance, the Mpemba effect can be conveniently characterized through the spectral decomposition of the master equation,
\begin{equation}
    \mathbf{P}_T(t)
    =
    \exp[M(T_b)t]\mathbf{P}_T(0)
    =
    \mathbf{P}^{ss}(T_b)
    +
    \sum_{i=2}^{6}
    a_i(T,T_b)
    e^{\lambda_i(T_b)t}
    \mathbf{V}_i(T_b),
    \label{Eq_eig_decomp}
\end{equation}
where $\mathbf{V}_i(T_b)$ are the right eigenvectors of $M(T_b)$, ordered according to their eigenvalues $\lambda_i(T_b)$. The coefficients are given by
\begin{equation}
    a_i(T,T_b)
    =
    \frac{\mathbf{W}_i(T_b)\cdot\mathbf{P}_T(0)}
    {\mathbf{W}_i(T_b)\cdot\mathbf{V}_i(T_b)},
    \label{Eq_a_i_def}
\end{equation}
where $\mathbf{W}_i(T_b)$ denote the corresponding left eigenvectors of $M(T_b)$. In general, when detailed balance is broken, the left and right eigenvectors are distinct.

\begin{figure}[htb!]
    \centering
    \includegraphics[width = \textwidth]{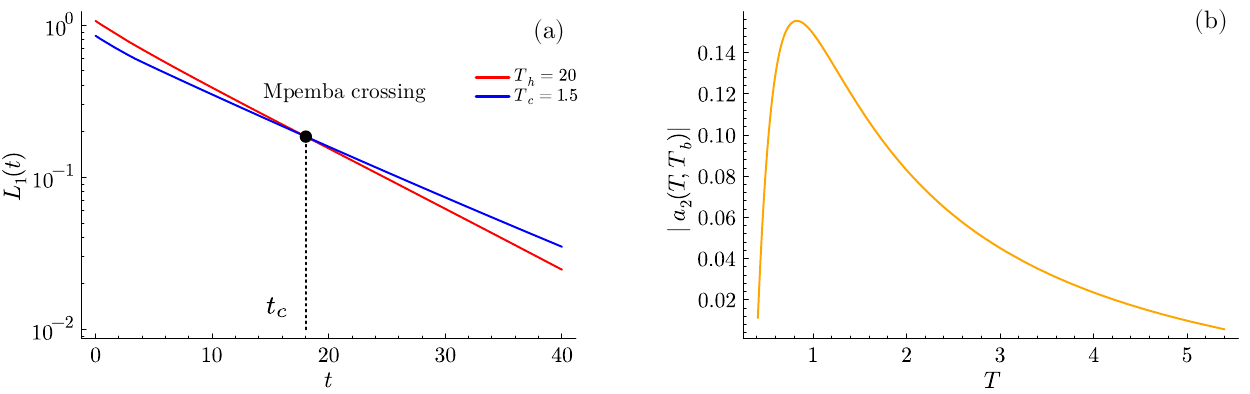}
    \caption{Examples illustrating equivalence of the spectral $a_2$ condition with the Mpemba effect. Panel (a):  Distance to steady state as a function of time for steady states prepared at $T_c = 1.5~\&~T_h = 20$, showing a Mpemba crossing at a time $t_c$. Parameters are $T_b = 0.4,~r = 0.1,~s = 2.5$. Panel (b): $|a_2(T,T_b)|$ vs $T$ with , where the $a_2$ curve displays non-monotonicity as a function of $T$ allowing for initial states that satisfy Eq.~(\ref{Eq_a_2_cond}). Other parameters same as Fig. \ref{Fig_mpemba_example}(a).} 
    \label{Fig_mpemba_example}
\end{figure}

Under equilibrium conditions, the long-time relaxation is governed by the slowest decaying non-stationary mode. Consequently, a non-monotonic dependence of the coefficient $a_2$ on the initial temperature provides both a necessary and sufficient criterion for the Mpemba effect. More precisely, the effect is present if
\begin{equation}
    a_2(T_h,T_b)
    <
    a_2(T_c,T_b).
    \label{Eq_a_2_cond}
\end{equation}
Thus, if $a_2(T,T_b)$ is not monotonically increasing as a function of $T$, we can choose temperatures $T_b<T_c<T_h$ in an interval where $a_2$ is decreasing such that Eq.~(\ref{Eq_a_2_cond}) is satisfied -- see Fig. \ref{Fig_mpemba_example}b.



\FloatBarrier

\section{Energy landscape}\label{section_energy_landscape}

As discussed in the Introduction, the free-energy landscape plays a central role in determining the Mpemba effect. It influences both the initial steady state from which the system is quenched and the subsequent relaxation dynamics toward the final steady state. Before investigating the Mpemba effect under different equilibrium and non-equilibrium conditions, we therefore first examine how the underlying free-energy landscape is modified by the control parameters of the model.

The three control parameters considered throughout this work are the bath temperature $T_b$, the external load force $F$, and the chemical concentrations. The latter are parameterized as
$
[ATP]=rs,\qquad [ADP]=s,
$
where $r$ denotes the ATP-to-ADP concentration ratio and $s$ sets the overall concentration scale. The phosphate concentration is then determined according to
$
[P]
=
c\frac{K_{eq}[ATP]}{[ADP]}
=
cr,
$
where the dimensionless parameter $c$ quantifies the degree of chemical non-equilibrium, with $c=1$ corresponding to chemical equilibrium.

\FloatBarrier
\begin{figure}[htb!]
    \centering
    \includegraphics[width = \textwidth]{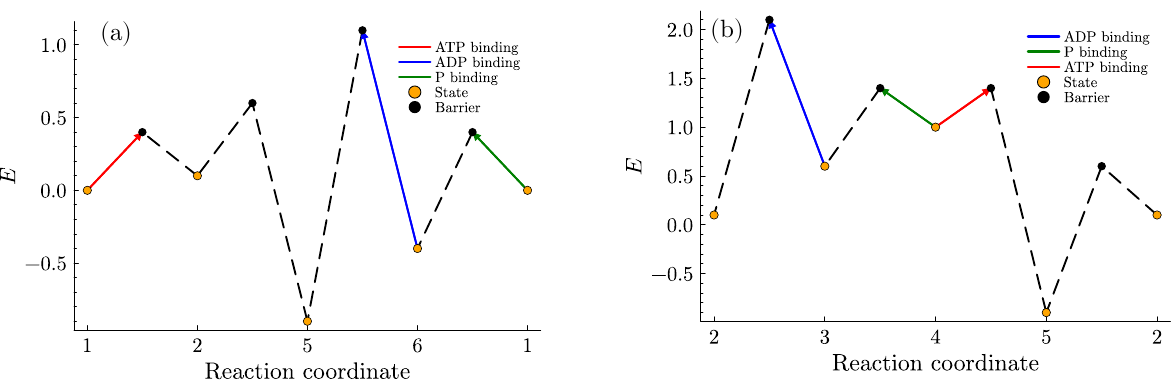}
    \caption{Energy landscape of the Kinesin walker that determines $\kappa_{ij}$ through Eq.~\ref{Eq_intrinsic_rate}. Solid lines denote transitions involving binding of chemical species, with color denoting type and arrow denoting direction of binding. Panel (a): Energy landscape for the forward cycle. Panel (b): Energy landscape for the backward cycle.} 
    \label{Fig_energy_landscape}
\end{figure}
\FloatBarrier

The resulting energy landscape possesses multiple metastable states, with states 2 and 6 corresponding to local minima, while state 5 forms the dominant energy minimum. The stability of these states can be systematically tuned through the model parameters. In particular, changing the chemical concentrations modifies the relative depths of the minima, thereby altering the degree of metastability of states such as 6 and, for suitable parameter values, generating additional metastable states, most notably states 1 and 4. In contrast, the external load acts primarily on the mechanical transition between states 2 and 5 and therefore changes the stability of the dominant minimum associated with state 5.

It is worth emphasizing that, unlike typical equilibrium systems, the steady-state distribution of this network does not necessarily become more homogeneous as the bath temperature is increased, since the chemical binding rates are limiting at high temperatures. The interplay between thermal activation, force-dependent transition rates, and chemical driving produces a nontrivial dependence of the steady-state occupations on temperature.

Throughout this work, we use the steady-state occupation probabilities as a practical measure of metastability: states with larger steady-state occupations are interpreted as being more stable, since they typically possess smaller effective escape rates. This correspondence is, however, only approximate because the occupation probability depends not only on the escape rates but also on the network connectivity. For example, states 2 and 5 each possess a larger coordination number than the remaining states, allowing them to sustain relatively high steady-state occupations even when their escape rates are comparatively large.

\section{Results}\label{section_results}
\subsection{Mpemba effect in equilibrium}\label{section_results_eq}
Although equilibrium Kinesin is of limited biological relevance, we begin with this case for two reasons. First, it provides the simplest setting in which the origin of the Mpemba effect can be understood without the additional complications introduced by broken detailed balance. Second, many of the physical insights gained from the equilibrium energy landscape extend naturally to the non-equilibrium regimes studied later. Equilibrium therefore serves both as a conceptual starting point and as a reference against which the effects of chemical and mechanical driving can be assessed.

Throughout this section, the equilibrium Mpemba phase diagrams are obtained using the spectral criterion in Sec.~\ref{section_mpemba_a2}. For each set of parameter values, we compute the coefficient $a_2(T,T_b)$ at 250 uniformly distributed temperatures satisfying $T_b+0.1<T<T_b+50$. The system is classified as exhibiting the Mpemba effect whenever these sampled values display a non-monotonic dependence on the initial temperature.

\FloatBarrier
\begin{figure}[htb!]
    \centering
    \includegraphics[width = \textwidth]{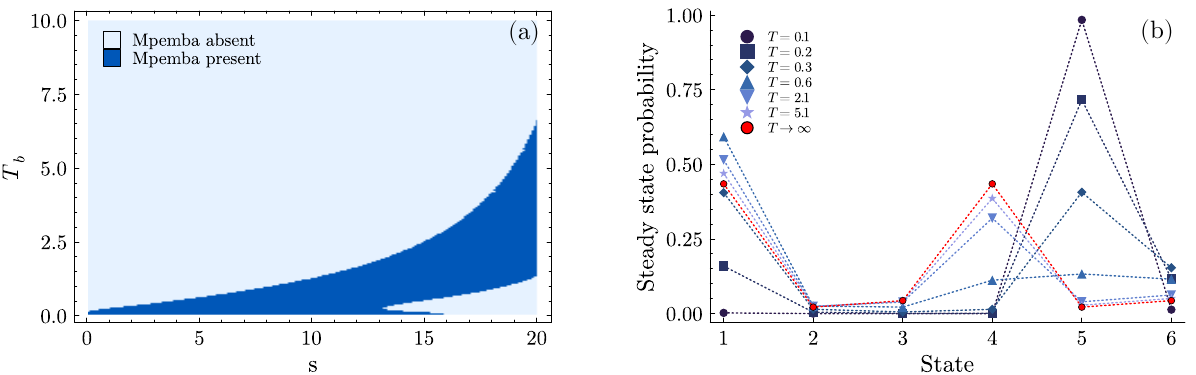}
    \caption{Panel (a): Mpemba phases in the plane of $T_b$ and $s$ for fixed $r = 0.1$ constructed from the spectral $a_2$-condition. Panel (b): Steady state occupation probabilities for different $T$'s with $s = 0.5,~ r = 0.1$  }
    \label{Fig_eq_phase_plot_ss}
\end{figure}
\FloatBarrier
Several qualitative features of the phase diagram shown in Fig.~\ref{Fig_eq_phase_plot_ss}(a) can be understood directly from the underlying energy landscape. We first consider the regime of small concentration scale $s$. At sufficiently large temperatures, the occupation probabilities of states 2 and 5 become small because the concentration-dependent transitions
$1\rightarrow2$, $3\rightarrow2$, $6\rightarrow5$, and $4\rightarrow5$
are suppressed relative to their reverse transitions. As a consequence, the steady-state probability shifts away from states 2 and 5 and becomes concentrated predominantly in states 1 and 4, as illustrated in Fig.~\ref{Fig_eq_phase_plot_ss}(b). For small values of the concentration ratio $r$, these two states become nearly equally populated in the high-temperature limit.

Since state 5 is the global energy minimum, the steady state at low bath temperatures is strongly localized around this state. Increasing the temperature gradually transfers probability from state 5 towards states 1 and 4. Importantly, the occupation of state 1 decreases while that of state 4 increases with temperature. Consequently, when the final bath temperature $T_b$ is sufficiently low for state 5 to remain the dominant minimum, relaxation from state 4 to state 5 is faster than from state 1 to state 5, giving rise to the Mpemba effect. As $T_b$ increases further, however, state 5 loses its dominant stability and this mechanism ceases to operate, leading to the disappearance of the Mpemba effect.

The opposite trend is observed upon increasing the concentration scale $s$. Larger values of $s$ enhance the incoming transitions into states 2 and 5 while leaving the corresponding outgoing transitions unchanged, thereby increasing their steady-state occupations. Since state 5 already dominates at low temperatures, state 2 emerges as an additional metastable state as the bath temperature increases, with the occupations of states 2 and 5 becoming comparable in the high-temperature limit. Consequently, for a fixed bath temperature, increasing $s$ progressively suppresses the Mpemba effect because the initial probability distribution becomes increasingly trapped in the metastable state 2.

\begin{figure}[htb!]
    \centering
    \includegraphics[width = \textwidth]{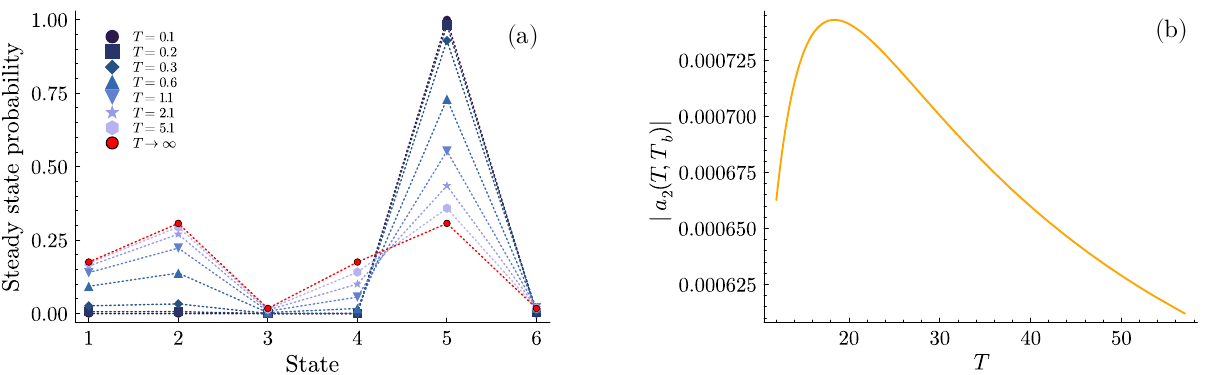}
    \caption{Example showcasing that metastability is not sufficient to predict the Mpemba effect. Panel (a): Steady state distributions for different $T$'s with $s = 17.5,~ r = 0.1$, where an increase in temperature induces metastability, suggesting that the Mpemba effect should be absent. Panel(b) Non monotonous $a_2$ curve  for $T_b = 7$ showing that the Mpema effect can nonethelss occur.}
    \label{Fig_mpemba_no_metastability}
\end{figure}

Although these trends can be qualitatively understood from the evolution of the energy landscape, this picture alone is insufficient to predict the precise location of the Mpemba phase boundaries. In particular, there exist parameter regimes, shown in Fig.~\ref{Fig_mpemba_no_metastability}, where increasing the temperature enhances metastability but the Mpemba effect nevertheless persists. This demonstrates that the presence of metastable states, while often correlated with anomalous relaxation, is not by itself a sufficient criterion for the occurrence of the Mpemba effect.

\FloatBarrier
\FloatBarrier

\begin{figure}[htb!]
    \centering
    \includegraphics[width = \textwidth]{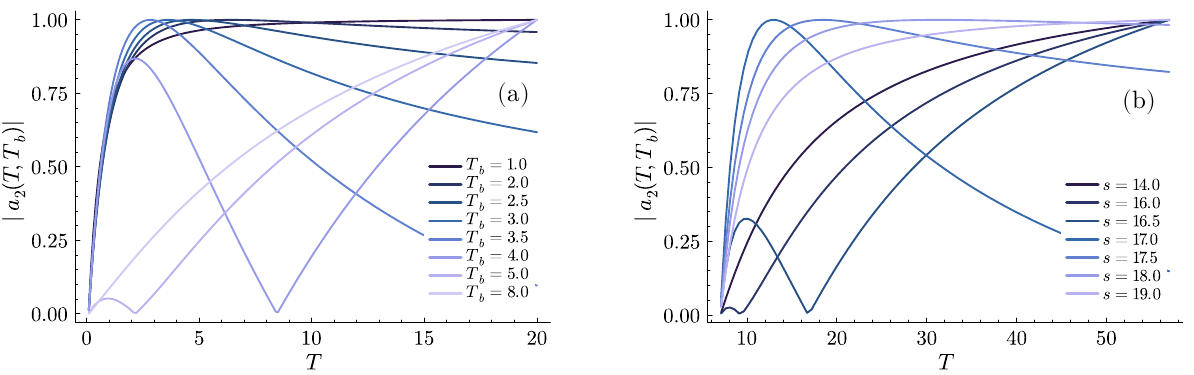}
    \caption{Variation in $a_2(T,T_b)$ with $T$ for different parameters showing the two distinct behaviors near Mpemba phase boundaries. Curves normalized to show shape and  $x$ axis re-centered at $T_b$ for each curve. Panel (a): $|a_2(T,T_b)|$ vs $T$ for different $T_b$ with $s = 15,~r =0.1$. Increasing $T_b$ corresponds to moving from right to left in the phase diagram shown in Fig.~\ref{Fig_eq_phase_plot_ss}. Panel (b): $|a_2(T,T_b)|$ vs $T$ for different $s$ with $T_b = 7,~r =0.1$. Increasing $s$ corresponds to moving from left to right in the phase plot in Fig.~\ref{Fig_eq_phase_plot_ss}.}
    \label{Fig_eq_a2_variation}
\end{figure}

Finally, the two boundaries of the Mpemba phase exhibit qualitatively different behaviors. As the right boundary is approached, the coefficient $a_2$ gradually evolves into a monotonically increasing function of the initial temperature. In contrast, near the left boundary the Mpemba effect survives only over a finite interval of initial temperatures, which continuously shrinks to zero upon approaching the boundary. This behavior is associated with a sign change of $a_2$, as illustrated in Figs.~\ref{Fig_eq_a2_variation}(a) and \ref{Fig_eq_a2_variation}(b). At the temperature for which $a_2(T)=0$, the slowest relaxation mode is absent from the dynamics, and the relaxation is instead governed by the next eigenmode, resulting in exponentially faster convergence to the steady state. This phenomenon is known as the \emph{strong Mpemba effect}~\cite{klichMpembaIndexAnomalous2019,kumarExponentiallyFasterCooling2020}.
\FloatBarrier
\FloatBarrier

\subsection{Mpemba effect out of equilibrium}\label{section_results_non_eq}
Although the equilibrium analysis provides valuable physical intuition, Kinesin operates under non-equilibrium conditions in vivo. We therefore next investigate how the Mpemba effect is modified when detailed balance is broken through either an external load force or chemical driving.

We first consider mechanical non-equilibrium by introducing a finite opposing load force while maintaining chemical equilibrium ($c=1$). Unless otherwise specified, the load-sharing parameter is fixed within the physical range $\theta\in[0,1]$, while the external force satisfies $F>0$. As discussed in Sec.~\ref{section_energy_landscape}, the load selectively modifies the mechanical transition rates, destabilizing state 5 while simultaneously stabilizing state 2. Consequently, depending on the underlying energy landscape, the applied force may either create or eliminate metastable states.

\begin{figure}[htb!]
    \centering
    \includegraphics[width = \textwidth]{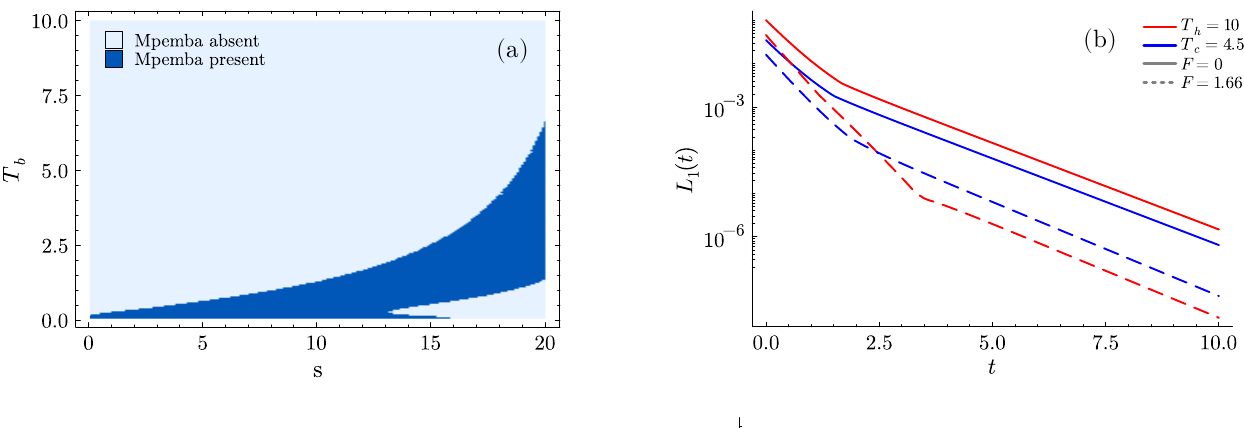}
    \caption{Plots showing the effect of adding external forcing. Panel (a) Mpemba phase diagram with $r = 0.1$ and $F = 1.66,~\theta = 0.6$, showing that a positive load largely supresses the Mpemba effect over a given parameter range. Panel (b): Example showcasing force induced Mpemba effect with $T_b = 3.5,~r = 0.1,~s = 18.5,~\theta = 0.6$. In the absence of a load, there is no crossing in distance to steady state while it is present on adding a load. } 
    \label{Fig_plots_forced}
\end{figure}

The influence of the load is naturally asymmetric with respect to its direction and primarily manifests itself as a stretching of the equilibrium phase diagram, causing the large-$s$ features to appear at progressively larger concentration scales. A representative example is shown in Fig.~\ref{Fig_plots_forced}(a). From the perspective of the transition rates, the applied load reduces the forward mechanical rate $R_{25}$, causing state 5 to lose its dominant stability at a lower bath temperature and allowing state 2 to emerge as a competing metastable state. Thus in the large $s$ regime where transitions between 2 and 5 determine long time relaxation, the Mpemba phase boundaries are pushed down for a given value of $s$. This provides another example of the intricate interplay between metastability and anomalous relaxation: although neither is a necessary nor a sufficient condition for the other, changes in the energy landscape strongly influence the occurrence of the Mpemba effect.

\FloatBarrier
\FloatBarrier

\subsection{Mpemba effect out of equilibrium with force quenching}\label{section_results_non_eq_force}
While the Mpemba effect is conventionally studied following temperature quenches, the same formalism can be applied to quenches of other control parameters, including the external load force~\cite{deguntherAnomalousRelaxationNonequilibrium2022}. In this protocol, the motor is initially prepared in the steady state corresponding to a load $F_i$, after which the force is suddenly changed to a new value $F_f$. Such rapid force changes are routinely employed in single-molecule experiments to probe the mechanochemical response of molecular motors. In the present context, the relaxation describes how rapidly the motor adapts to an abrupt change in the applied load. The spectral analysis proceeds exactly as for temperature quenches, with the coefficient $a_2(F)$ replacing $a_2(T)$.

\begin{figure}[htb!]
    \centering
    \includegraphics[width = \textwidth]{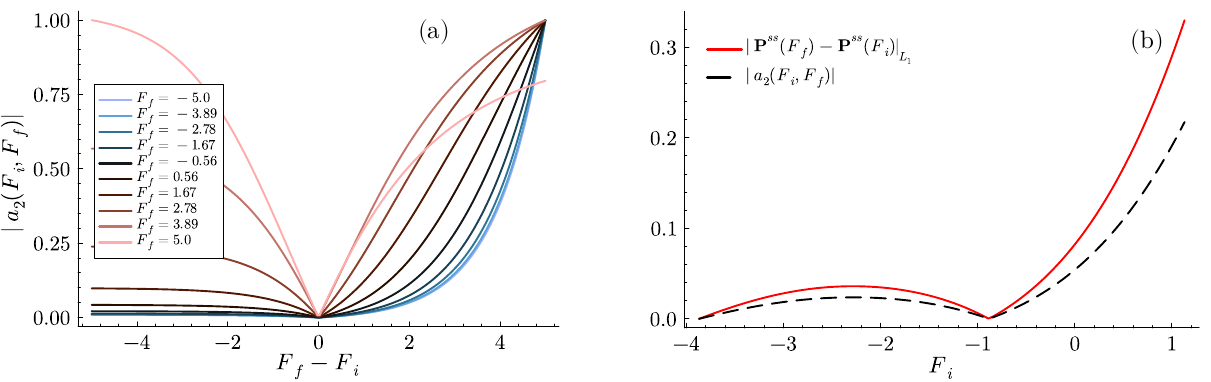}
    \caption{Examples of behavior seen after a force quench. Panel(a) Example of $a_2(F_i,F_f)$ which determines long time relaxation behavior after a sudden change of load from $F_i$ to $F_f$. Paramteres are $T_b = 0.2,~s = 18.5,~r = 0.1,~\theta = 0.6$. The $y$ axis is normalized so that the curve shapes can be compared. Panel(b): An example showing the variation in $a_2(F_i,F_f)$ and initial distance to steady state overlapping with $F_f = -3.87,~ T_b = 0.1,~ s = 0.1,~ r = 0.1,~ \theta = -1.2$.}
    \label{Fig_force_quench}
\end{figure}

Interestingly, despite an extensive exploration of parameter space, we do not observe a force-induced Mpemba effect. A representative dependence of $a_2(F)$ is shown in Fig.~\ref{Fig_force_quench}(a). Although certain parameter regimes possess complex eigenvalues, the corresponding imaginary parts are too small for oscillatory relaxation to become visible before the distance to the steady state falls below numerical precision. Another noteworthy feature is illustrated in Fig.~\ref{Fig_force_quench}(b). For sufficiently large values of $\theta$, the change in the steady-state distribution induced by varying the load becomes nearly parallel to the slowest relaxation mode. Consequently, any non-monotonicity in $a_2(F)$ is accompanied by a similar non-monotonicity in the initial distance from the steady state, thereby violating the basic requirement for observing the Mpemba effect. The figure also demonstrates that the steady-state distribution is not uniquely determined by the applied load.

\FloatBarrier
\FloatBarrier

We finally consider chemical non-equilibrium by allowing the ATP hydrolysis reaction to operate away from equilibrium. The chemical driving is controlled by the parameter $c$, which determines the phosphate concentration while the ATP and ADP concentrations are held fixed. Varying $c$ effectively modifies the chemically activated transition barriers shown in Fig.~\ref{Fig_energy_landscape}. 

Similar to the effect of an external load, chemical driving primarily stretches or compresses the equilibrium phase diagram without qualitatively altering its overall structure. Furthermore, although detailed balance is broken, we do not observe oscillatory relaxation over the parameter range investigated.

\FloatBarrier
\begin{figure}[htb!]
    \centering
    \includegraphics[width = \textwidth]{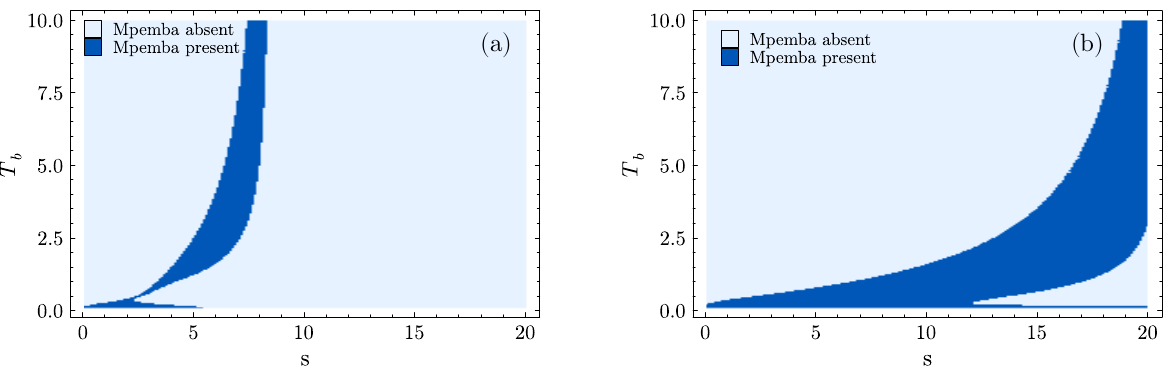}
    \caption{Effect of $c$ on the Mpemba phase diagram with all other parameters as in Fig. 5 (a). Panel (a): $T_b$ vs $s$ phase diagram for $c = 5.5$ where the Mpemba effect is largely suppressed over. Panel (b): $T_b$ vs $s$ phase diagram for $c = 0.1$, where the Mpemba effect is seen over wider parameter ranges compared to equilibrium.} 
    \label{Fig_plots_chemically_driven}
\end{figure}
\FloatBarrier

\subsection{Velocity or current as a probe of the Mpemba effect}\label{section_results_current}
The Mpemba effect is traditionally characterized by monitoring the relaxation of the full probability distribution toward its steady state. While this provides a natural theoretical framework, the occupation probabilities of the microscopic states are generally difficult to access in experiments. For molecular motors, however, experimentally measurable observables such as the motor velocity are routinely monitored with high precision. An important question is therefore whether anomalous relaxation can be inferred directly from the dynamics of such observables. In the context of the Mpemba effect in open quantum systems, ``good'' observables have been proposed as a method to bypass the requirement for state tomographies \cite{baguiDetectionMpembaEffect2026}. In this section, we show that the relaxation of the motor velocity faithfully reflects the Mpemba effect, thereby providing a direct route for its experimental observation.

Single-molecule tracking has long been the primary experimental technique for studying the stepping mechanism of Kinesin. In these experiments, a micron-sized bead is attached to the motor and manipulated using an optical trap, which exerts an approximately Hookean load on the motor. By varying both the trap strength and the chemical environment, one can probe the chemomechanical properties of Kinesin motion with nanometer precision, allowing individual 8~nm steps to be resolved~\cite{blockBeadMovementSingle1990,svobodaForceVelocityMeasured1994,carterMechanicsKinesinStep2005}. The resulting trajectories are routinely used to determine the step-averaged velocity, which serves as one of the principal observables for validating kinetic models of molecular motors. This naturally motivates the motor velocity as an experimentally accessible probe of anomalous relaxation.

\begin{figure}[htb!]
    \centering
    \includegraphics[width = \textwidth]{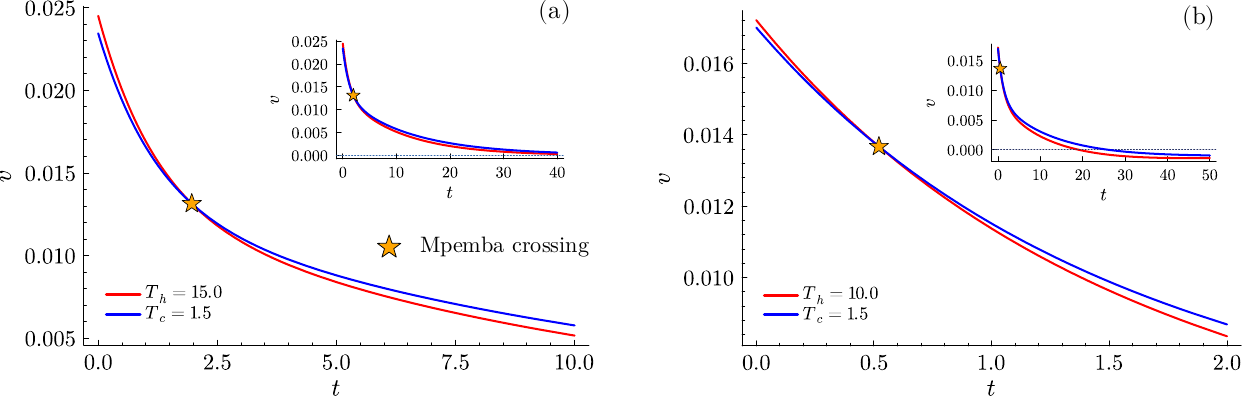}
    \caption{Examples showing crossing over of the probability current $J_{25}$, corresponding to the walker velocity $v$ during relaxation. Panel (a): Equilibrium scenario where the current decays to zero with $r = 0.1,~s = 2.5,~T_b = 0.4$. Panel(b): Out of equilibrium case where the steady state current is non-zero. Parameters are $c = 1.0,~F = 0.4,~\theta = 0.1,~r = 0.1,~s = 1.5,~T_b = 0.4$. Insets indicate saturation of the velocities in their corresponding equilibrium and non-equilibrium steady state values respectively.} 
    \label{Fig_current_mpemba}
\end{figure}

Following a quench, the probability currents also relax exponentially towards their steady-state values as shown in Fig.~\ref{Fig_current_mpemba}. While these currents vanish at equilibrium, they generally remain finite in non-equilibrium steady states. Recalling that the probability current across the transition $i\rightarrow j$ is

\begin{equation}
J_{ij}
=
\mathbf{P}_iR_{ij}
-
\mathbf{P}_jR_{ji},
\label{Eq_Jij}
\end{equation}
where its relaxation is given by
\begin{align}
J_{ij}(t)-J_{ij}^{ss}
&=
[\mathbf{P}_i(t)-\mathbf{P}_i^{ss}]R_{ij}
-
[\mathbf{P}_j(t)-\mathbf{P}_j^{ss}]R_{ji}
\label{Eq_Jij_minus_Jss}
\\
&=
\sum_{n=2}^{6}
a_n(T,T_b)
e^{\lambda_n(T_b)t}
\underbrace{
\left[
V_{n,i}(T_b)R_{ij}(T_b)
-
V_{n,j}(T_b)R_{ji}(T_b)
\right]
}_{C_{n,ij}(T_b)},
\label{Eq_Jij_eig_decomp}
\end{align}
where the second equality follows from inserting the eigenmode expansion of $\mathbf{P}(t)$ [Eq.~(\ref{Eq_eig_decomp})], and $V_{n,k}$ denotes the $k$-th component of the right eigenvector $\mathbf{V}_n$. Consequently, provided the projection coefficient $C_{2,ij}$ is non-zero, the long-time relaxation of the current is governed by exactly the same slow mode that determines the relaxation of the probability distribution. The current therefore exhibits the Mpemba effect whenever the $a_2$ criterion is satisfied.

\FloatBarrier
\FloatBarrier

For Kinesin, the mechanically relevant current $J_{25}$ is precisely the motor velocity. We find that the corresponding projection coefficient $C_{2,25}$ remains non-zero throughout the parameter ranges investigated, implying that the velocity faithfully reproduces the anomalous relaxation of the full stochastic dynamics. This correspondence continues to hold even out of equilibrium, provided oscillatory relaxation is absent, as illustrated in Fig.~\ref{Fig_current_mpemba}(b).

Our results therefore demonstrate that the Mpemba effect can be detected without reconstructing the full probability distribution, but instead through direct measurements of the motor velocity, providing a realistic experimental route for observing anomalous relaxation in chemomechanical systems.

\section{Conclusion}\label{section_conclusion}
In this work, we investigated anomalous relaxation in the six-state chemomechanical network model of the Kinesin molecular motor. We considered a biologically relevant kinetic network following Lipowsky et al and examined how anomalous relaxation evolves under both equilibrium and non-equilibrium conditions. In particular, we independently explored the effects of mechanical driving through an external load force and chemical driving through non-equilibrium ATP hydrolysis.

We showed that many of the qualitative features of the Mpemba phase diagram can be understood from the evolution of the underlying free-energy landscape. Changes in the metastability of the kinetic states provide valuable physical intuition for the emergence or disappearance of the Mpemba effect. At the same time, our results demonstrate that metastability alone is not a sufficient condition for anomalous relaxation, emphasizing that the complete spectral structure of the stochastic dynamics ultimately determines the Mpemba phase boundaries. We further identified the distinct mechanisms by which the Mpemba effect disappears and related one of them to the onset of the strong Mpemba effect.

Breaking detailed balance through either mechanical or chemical driving preserves the overall phenomenology of anomalous relaxation, primarily modifying the location and extent of the Mpemba phase rather than its qualitative structure. Furthermore, although non-equilibrium dynamics may in principle generate oscillatory relaxation through complex eigenmodes, such behavior was not observed over the physically relevant parameter ranges explored in this work. We also investigated quenches of the external load force itself and found no evidence of a force-induced Mpemba effect, highlighting an important distinction between temperature and force quenches in chemomechanical systems.

Another significant outcome of this work is the identification of the motor velocity as an experimentally accessible signature of the Mpemba effect. Rather than reconstructing the full probability distribution of the microscopic states, which is generally inaccessible in experiments, one can infer anomalous relaxation directly from the relaxation of the measurable probability current. Since motor velocity is routinely measured with high precision in single-molecule experiments~\cite{blockBeadMovementSingle1990,svobodaForceVelocityMeasured1994,carterMechanicsKinesinStep2005}, this provides a realistic route for experimentally probing the Mpemba effect in molecular motors and, more broadly, in non-equilibrium biochemical networks.

More generally, we hope that this work stimulates the exploration of anomalous relaxation in living systems. Many biological processes are naturally described by finite-state stochastic networks operating far from equilibrium\cite{geStochasticTheoryNonequilibrium2012}, including molecular motors\cite{kolomeiskyMotorProteinsMolecular2013}, enzyme catalytic cycles\cite{qianPhosphorylationEnergyHypothesis2007}, ion-channel gating, gene regulatory networks, biochemical proofreading, and molecular chaperones \cite{beardChemicalBiophysicsQuantitative2008}. It would therefore be interesting to investigate whether signatures of the Mpemba effect arise in other experimentally established kinetic models, such as the chemomechanical models of dynein and myosin\cite{schnitzerStatisticalKineticsProcessive1995}, kinetic proofreading networks\cite{hopfieldKineticProofreadingNew1974,ninioKineticAmplificationEnzyme1975}, or models of protein folding and conformational switching\cite{hyeonStructuralPerspectiveDynamics2011}. More broadly, extending the present framework to non-Markovian dynamics, active biochemical processes, and interacting molecular networks may help clarify whether anomalous relaxation plays a functional role in biological adaptation, information processing, or energy transduction.  

\section{Acknowledgements}
KC acknowledges IMSc for technical support, research facilities and funding through the visiting students programme. 
AP acknowledges research funding under the scheme
ANRF/ARGM/2025/001623 from ANRF, India and research support from the Department of Science and Technology, India, SERB Start-up Research Grant Number SRG/2022/000080. AP also acknowledges the International Research Project (IRP) titled ``Classical and quantum dynamics in out of equilibrium systems'' by CNRS, France. Finally, we gratefully acknowledge research support from the Department of Atomic Energy, Government of India via Soft Matter Apex projects.

\printbibliography
\end{document}